\documentclass[twocolumn,
pra
,superscriptaddress
,floatfix
,aps
,10pt
,showpacs]{revtex4-1}

\usepackage{graphicx}
\usepackage{amssymb}
\usepackage{amsmath}
\usepackage{multirow}
\usepackage{array,tabularx}
\usepackage{xcolor}

\usepackage[colorlinks,urlcolor=blue,citecolor=blue,linkcolor=blue]{hyperref}

\DeclareMathOperator*{\Simiq}{\simeq}

\DeclareMathOperator*{\sump}{\bar{\sum}}

\newcommand{\vect}[1]{{\mathbf #1}}
\newcommand{\Frac}[2]{\displaystyle\frac{#1}{#2}}

\newcommand{\bra}[1]{\langle\left.{#1}\right|}
\newcommand{\ket}[1]{\left|{#1}\right.\rangle}

\usepackage{float}
\usepackage{upgreek}
\usepackage{makeidx}
\DeclareGraphicsExtensions{.png,.jpg,.pdf,.mps,.gif,.bmp,.eps}
\usepackage{physics}

\renewcommand\vec{\mathbf}
\renewcommand\zeta{\xi} 

\begin{document}


\title{Early-time dynamics of Bose gases quenched into the strongly interacting regime}

\author{A. Mu\~noz de las Heras}
\email{alberto.munnozd@estudiante.uam.es}
\affiliation{Departamento de F\'isica Te\'orica de la Materia
  Condensada \& Condensed Matter Physics Center (IFIMAC), Universidad
  Aut\'onoma de Madrid, Madrid 28049, Spain}
  
\author{M. M. Parish}
\affiliation{School of Physics \& Astronomy, Monash University, Victoria 3800, Australia}  

\author{F. M. Marchetti}
\email{francesca.marchetti@uam.es}
\affiliation{Departamento de F\'isica Te\'orica de la Materia
  Condensada \& Condensed Matter Physics Center (IFIMAC), Universidad
  Aut\'onoma de Madrid, Madrid 28049, Spain}

\date{November 29, 2018}

\begin{abstract}
  We study the early-time dynamics of a degenerate Bose gas after a
  sudden quench of the interaction strength, starting from a weakly
  interacting gas. By making use of a time-dependent generalization of
  the Nozi\`eres-Saint-James variational formalism, we describe the
  crossover of the early-time dynamics from shallow to deep
  interaction quenches.
  We analyze the coherent oscillations that characterize both the
  density of excited states and the Tan's contact as a function of the
  final scattering length. For shallow quenches, the oscillatory
  behaviour is negligible and the dynamics is universally governed by
  the healing length and the mean-field interaction energy. By
  increasing the final scattering length to intermediate values, we
  reveal a universal regime where the period of the coherent
  atom-molecule oscillations is set by the molecule binding energy.
  For the largest scattering lengths we can numerically simulate in
  the unitary regime, we find a universal scaling behaviour of the
  typical growth time of the momentum distribution in agreement with
  recent experimental observations [C. Eigen \emph{et al.}, Nature
  \textbf{563}, 221 (2018)].
\end{abstract}


\maketitle

\section{Introduction}
\label{sec:Introduction}
The possibility of tuning the effective interatomic interaction in
ultracold atomic gases via a magnetic-field Feshbach
resonance~\cite{FeshbachReviewChin} opens up novel perspectives in the
exploration of quantum many-body phenomena and strongly correlated
behavior.
In particular, it is now possible to experimentally investigate
strongly interacting Bose gases near the unitary limit, where the
$s$-wave scattering length diverges, $a \to
\infty$~\cite{Fletcher_PRL_2013,Rem_PRL2013,Makotyn,Eismann_PRX_2016,Klauss_PRL_2017,Lopes_PRL_2017,Fletcher,UniversalScalingLaws_Eigen_2017,UniversalPrethermalDynamicsHadzibabic}.
Such a system is inherently unstable due to strong three-body losses
to lower lying states, and thus experiments must access the strongly
interacting regime by starting with a weakly interacting Bose gas and
then rapidly ramping the external magnetic field to the desired
scattering length. However, rather than being a complication, this
provides a unique opportunity to study the dynamics of a quantum
many-body system driven out of equilibrium in a controlled manner.
For instance, for shallow quenches within the weakly interacting
regime, it has been possible to reconstruct the spectrum of collective
excitations of the degenerate gas from the characteristic oscillations
of the density-density correlation function~\cite{Hung_2013}, which is
the analogue of cosmological Sakharov oscillations.

For the case of deep quenches into the unitary regime, there are
fundamental questions regarding the role and nature of three-body
losses in the evolution of the Bose
gas~\cite{Fletcher_PRL_2013,Rem_PRL2013,Klauss_PRL_2017,Eismann_PRX_2016,UniversalScalingLaws_Eigen_2017}.
A pioneering experiment~\cite{Makotyn} has shown that the momentum
distribution of a degenerate gas can saturate to a prethermal
steady-state distribution, thus suggesting that the early-time
dynamics after a deep quench is unaffected by three-body
recombination.
Most recently, C. Eigen \emph{et
  al.}~\cite{UniversalPrethermalDynamicsHadzibabic} have observed a
universal prethermal post-quench dynamics of the degenerate gas at
unitarity. Here, the authors have found scaling laws of the
momentum-distribution growth time versus momentum, which are universal
when expressed in terms of density scales. These results indicate
that, in the post-quench dynamics at very short times, the
quasi-particle excitations of the unitary gas are qualitatively
similar to the Bogoliubov modes. As we show in this paper, this
behaviour can be described without including three-body scattering or
losses.

The observed universal dynamics of the unitary Bose gas is consistent
with the scale invariance one expects when the scattering length
diverges and there is no interaction scale in the problem. This would
appear to contradict the existence of Efimov
trimers~\cite{Efimov_1970}, which introduces another length scale in
the problem and thus breaks the scale invariance of the degenerate
Bose gas~\cite{Braaten2006}. However, experiments on the thermal
gas~\cite{Fletcher} have shown that the three-body contact
$C_3$~\cite{HudsonSmith_2014} (which provides a measure of the
three-body correlations connected with Efimov trimers) is negligible
in an early time window after a quench into the unitary
regime. Therefore, this suggests that Efimov physics is only important
at later times in the dynamics, which is reasonable given that we
expect short times to be dominated by short-distance, two-body
scattering. Furthermore, the typical density range in many experiments
may be insensitive to the Efimov
scale~\cite{UniversalScalingLaws_Eigen_2017}.

In general, the quench dynamics of Bose gases near unitarity is a
challenge to describe
theoretically~\cite{YinRadzihovsky2013,Sykes,Rancon_Levin_PRA_2014,KainLing,CorsonBohn,Kira_2015,Ancilotto_2015,PostquenchRadzihovsky}
since there is no small interaction parameter that allows a
perturbative expansion. However, in the early stages of the dynamics
after a quench from weak interactions, we expect three-body
correlations and losses to be negligible, as discussed
above. Therefore, a theory involving only pairwise excitations out of
the condensate should be reasonable at short times. To this end, we
employ here a time-dependent generalization of the Nozi\`eres-Saint
James variational formalism~\cite{NSJ}, similarly to the approach in
Refs.~\cite{Sykes,CorsonBohn}. This also allows us to capture the
effect of the molecular two-body bound state, which is present on the
repulsive side of the resonance, $a>0$.

We go beyond previous work and analyze the crossover from shallow to
deep quenches using the variational formalism. For shallow quenches
the early-time dynamics is
integrable~\cite{NatuMuellerCorrelationsBogoliubov,Hung_2013,PrethermalisationThermalisationCrossoverIacopo}
and one can make use of a time-dependent Bogoliubov
approximation. However, in order to describe the crossover to deep
quenches, we have to resort to a numerical analysis of the coupled
dynamical equations for the condensate density and the excited state
distribution, which, in our modelling, includes the condensate
depletion as well as the correlations between non-condensed atoms.

We describe the coherent oscillations developed by the number of
excited particles as well as by the Tan's contact and, up to
intermediate values of the final scattering length $a_f$, we disclose
a universal regime where the oscillation period is constant and
determined by the inverse molecule binding energy only. We discuss the
optimal values of $a_f$ at which these atom-molecule coherent
oscillations should be detectable in current experiments.

For the largest scattering lengths we can numerically simulate in the
unitary regime, we find that the coherent oscillations persist and are
now due to the condensate interaction with the medium. While these
oscillations appear strongly damped in current
experiments~\cite{UniversalPrethermalDynamicsHadzibabic}, our model
can correctly reproduce the universal scaling behaviour of the typical
growth time of the momentum distribution found by
Ref.~\cite{UniversalPrethermalDynamicsHadzibabic}, suggesting that
higher-order damping mechanisms do not affect the postquench behaviour
at very-early times, when a steady state regime is not yet reached.

The paper is organized as follows: In Sec.~\ref{sec:model} we
introduce the variational Ansatz for describing the quench dynamics
and, in Sec.~\ref{sec:eqmot}, we derive the coupled equations of
motion for the condensate density and excited state distribution. We
summarize in Sec.~~\ref{sec:param} the characteristic system
parameters as well as the relevant length and time scales in the
crossover from shallow to deep quenches. In Sec.~\ref{sec:resul}, we
present the numerical results for the non-condensed fraction
(Sec.~\ref{sec:nonco}) and the Tan's contact (Sec.~\ref{sec:Tans}),
while in Sec.~\ref{sec:UniPreDy} we explore the universal scaling laws
associated with the very-early-time dynamics of the gas momentum
distribution in the unitary regime. Conclusions and perspectives of
our work are summarized in Sec.~\ref{sec:conc}.

%
\section{Model}
\label{sec:model}
We start by considering the Hamiltonian describing a homogeneous gas
of $N$ interacting bosons in a three-dimensional (3D) volume $V$
(henceforth we fix $\hbar = 1$)
\begin{equation}
  \hat{H} = \sum_{\vec{k}} \epsilon_{\vec{k}}^{}
  \hat{a}_{\vect{k}}^{\dagger} \hat{a}_{\vect{k}}^{} +
  \frac{U_\Lambda}{2V}\sum_{\vec{k}_1 , \vec{k}_2 , \vec{q} }
  \hat{a}_{\vec{k}_1+\vec{q}}^{\dagger}
  \hat{a}_{\vec{k}_2-\vec{q}}^{\dagger} \hat{a}_{\vec{k}_2}^{}
  \hat{a}_{\vec{k}_1}^{} \; ,
\label{eq:hamil}
\end{equation}
where $\hat{a}_{\vec{k}}^{\dagger}$ ($\hat{a}_{\vec{k}}^{}$) creates
(annihilates) a boson with momentum $\vec{k}$ and mass $m$
($\epsilon_{\vec{k}}^{} = |\vec{k}|^2/2 m \equiv k^2/2m$).  Close to a
Feshbach resonance, atom-atom interactions can be modelled via a
short-range pseudo-potential, which, in momentum space, is constant
with strength $U_\Lambda$ up to a momentum cutoff $\Lambda$. The
coupling constant and cutoff are related to the $s$-wave scattering
length $a$ through the $T$-matrix renormalization
process~\cite{FeshbachReviewChin,FeshbachReviewGurarieRadzihovsky}:
\begin{equation}
  \frac{m}{4\pi a} = \frac{1}{U_\Lambda} +
  \frac{1}{V}\sum_{\vec{k}}^{k<\Lambda} \frac{1}{2\epsilon_\vec{k}}
  =\frac{1}{U_\Lambda} + \frac{m\Lambda}{2\pi^2}\; .
\label{eq:U_Lambda}
\end{equation}
The (inverse) cutoff $\Lambda^{-1}$ represents the range of the
interaction potential, which is assumed to be much smaller than all
other length scales in the problem. In the limit $\Lambda \to \infty$,
the contact potential admits, on the repulsive side of the resonance
$a>0$, a single molecular bound state with
energy~\cite{FeshbachReviewGurarieRadzihovsky}:
\begin{equation}
  E_{\text{B}}=-\Frac{1}{ma^2} \; .
\label{eq:bound}
\end{equation}
We have checked that our results are converged with respect to the
cutoff $\Lambda$.

In order to separate the contribution of the condensed state $\vect{k}
= \vect{0}$ from that of the excited states $\vect{k} \ne \vect{0}$,
it is useful to rewrite the Hamiltonian~\eqref{eq:hamil} by
substituting $\hat{a}_{\vec{k}}^{} \rightarrow
\hat{a}_{0}^{}\delta_{\vec{k}\, ,\vect{0}}^{} + \hat{a}_{\vec{k}\neq
  \vect{0}}^{}$. One thus obtains the following contributions to the
Hamiltonian (we henceforth use the notation $\hat{a}_{\vec{k}}^{}$ for
$\hat{a}_{\vec{k}\neq \vect{0}}^{}$):
\begin{align}
\label{eq:decom}
  \hat{H} &= \sum_{\vec{k}} \epsilon_{\vec{k}}^{}
  \hat{a}_{\vect{k}}^{\dagger} \hat{a}_{\vect{k}}^{} +
  \frac{U_{\Lambda}}{2V}\hat{a}_{0}^{\dagger}\hat{a}_{0}^{\dagger}\hat{a}_{0}^{}\hat{a}_{0}^{}
  + \hat{H}^{}_2 + \hat{H}^{}_3 + \hat{H}^{}_4 \\
  \hat{H}^{}_2 &=\frac{U_{\Lambda}}{2V}\sum_{\vec{k}}\left(
  \hat{a}_{\vec{k}}^{\dagger}\hat{a}_{-\vec{k}}^{\dagger}\hat{a}_{0}^{}\hat{a}_{0}^{}
  + 2 \hat{a}_{\vec{k}}^{\dagger}\hat{a}_{\vec{k}}^{}
  \hat{a}_{0}^{\dagger} \hat{a}_{0}^{} + \text{h.c.}  \right)\\
  \hat{H}^{}_3 &=\frac{U_{\Lambda}}{V}\sum_{\vec{k}, \vec{q}}\left(
  \hat{a}_{\vec{k}-\vec{q}}^{\dagger}
  \hat{a}_{\vec{q}}^{\dagger}\hat{a}_{\vec{k}}^{}\hat{a}_{0}^{} +
  \text{h.c.}  \right)\\
\hat{H}^{}_4 &= \frac{U_\Lambda}{2V}\sum_{\vec{k}_1 , \vec{k}_2 ,
  \vec{q} } \hat{a}_{\vec{k}_1+\vec{q}}^{\dagger}
\hat{a}_{\vec{k}_2-\vec{q}}^{\dagger} \hat{a}_{\vec{k}_2}^{}
\hat{a}_{\vec{k}_1}^{} \; .
\end{align}

In the following, we focus on zero temperature and consider the
early-time dynamics of the gas after a quench from an initial
scattering length $a_i$ to the final value $a_f$.  We will not carry
out any approximation at the level of the
Hamiltonian~\eqref{eq:decom}, unlike what is done in the
time-dependent self-consistent Bogoliubov
approximation~\cite{PostquenchRadzihovsky}.
Instead, we will use the time-dependent generalization of the
Nozi\`eres-Saint-James variational Ansatz~\cite{NSJ,Zhou_PRL_2009}:
\begin{equation}
  \ket{\psi (t)} = \frac{1}{\mathcal{N}(t)} 
  e^{\sqrt{V} c^{}_{0}(t) \hat{a}^{\dagger}_{0} + 
    \frac{1}{2} \sump_{\vec{k}} g^{}_{\vec{k}}(t) \hat{a}^{\dagger}_{\vec{k}}
    \hat{a}^{\dagger}_{-\vec{k}} } \ket{0} \; .
\label{eq:Ansatz}
\end{equation}
Here, the normalization constant $\mathcal{N}(t)$ ensures that
$\bra{\psi (t)}\ket{\psi (t)}=1$.
The complex variational parameters $c^{}_{0}(t)$ and
$g^{}_{\vec{k}}(t)$ are related to the momentum occupation numbers,
\begin{align}
  N^{}_{0}(t) &=|\langle \hat{a}^{}_{0} \rangle|^2 =
  V|c_0(t)|^2\\
  N_{\vec{k}}(t) & = \langle\hat{a}^{\dagger}_{\vec{k}}
  \hat{a}^{}_{\vec{k}}\rangle = \frac{|g_{\vec{k}}(t)|^2}{1 -
    |g_{\vec{k}}(t)|^2} \; ,
\end{align}
as well as to the pairing term:
\begin{equation}
  x^{}_{\vec{k}}(t) = \langle
  \hat{a}^{}_{\vec{k}}\hat{a}^{}_{-\vec{k}} \rangle =
  \frac{g_{\vec{k}}(t)}{1-|g_{\vec{k}}(t)|^2} \; ,
\end{equation}
where $\langle \dots \rangle = \bra{\psi(t)} \dots
\ket{\psi(t)}$. Note that $N_{\vec{k}}(t)$ and $x_\vec{k}(t)$ are not
independent functions, since they are related by the constraint
$|x_\vec{k}(t)|^2=N_\vec{k}(t)\left[N_\vec{k}(t)+1\right]$.
Also note that $g^{}_{-\vec{k}}(t) =g^{}_{\vec{k}}(t)$, while the factor
$1/2$ in the momentum sum in Eq.~\eqref{eq:Ansatz} avoids
double-counting. We use the notation $\sump_{\vec{k}}= \sum^{k<
  \Lambda}_{\vec{k}\neq \vect{0}}$.

The Ansatz $\ket{\psi (t)}$ describes the $\vec{k}=\vec{0}$ condensed
state as a coherent state, while particles at finite momentum $\vec{k}
\ne \vec{0}$ are excited out of the condensate in pairs only. For
shallow quenches ($n a_{i,f}^3 \ll 1$), the condensate depletion is
small, such that one can assume that the condensate density is time
independent, $|c_0(t)|^2 \simeq n$. In this case, the
Ansatz~\eqref{eq:Ansatz} is a controlled approximation at an early
stage of the dynamics, where Beliaev-Landau scattering processes
involving three particles can be safely
neglected~\cite{PrethermalisationThermalisationCrossoverIacopo}. Here,
one can neglect the contributions of $\hat{H}^{}_3$ and $\hat{H}^{}_4$
to the Hamiltonian~\eqref{eq:decom} and show that the dynamics becomes
integrable (see App.~\ref{app:shall}), which recovers the results
obtained by Refs.~\cite{NatuMuellerCorrelationsBogoliubov,Hung_2013}
within a time-dependent Bogoliubov approximation.

The same Ansatz~\eqref{eq:Ansatz} has previously been employed by
Refs.~\cite{Sykes,CorsonBohn} to describe the quench dynamics of a
Bose gas into the strongly interacting regime.  Assuming a weakly
interacting initial state ($n a_{i}^3 \ll 1$), we will use the
Ansatz~\eqref{eq:Ansatz} to study the crossover of the early-time
dynamics from shallow to deep quenches, where $n a_{f}^3 \ll 1$ and $n
a_f^3 \gtrsim 1$, respectively. In general, the dynamics of the
condensate depletion cannot be neglected and we include the time
dependence of the variational parameter $c_0(t)$ in a similar way to
the self-consistent Bogoliubov approximation considered by
Ref.~\cite{PostquenchRadzihovsky}.
Further, as explained later, we include all contributions arising from
$\langle \hat{H}^{}_{4} \rangle$ which describes the correlations
between non-condensed atoms. This, together with the renormalization
of the interaction coupling constant~\eqref{eq:U_Lambda}, allows one
to describe the atom-molecule coherent dynamics, as already discussed
in Ref.~\cite{CorsonBohn}.
However, as $\langle \hat{H}^{}_{3} \rangle = 0$, the Ansatz for
$\ket{\psi (t)}$ does not admit neither Beliaev decay nor Landau
damping terms that may be responsible for the loss of the
atom-molecule coherence.

For weak interactions,
Ref.~\cite{PrethermalisationThermalisationCrossoverIacopo} has shown
that there is a separation of time scales between short-time dynamics
dominated by pair-wise excitations and longer-time dynamics requiring
the inclusion of higher-order excitation terms such as
$\hat{H}^{}_{3}$. However, it remains an open question how this
separation of time scales evolves as we approach unitarity. For the
case of an impurity atom immersed in a quantum medium, it can be
formally shown that the dynamics following a quench of the
impurity-medium interactions to the unitary regime is dominated by
two-body correlations at times less than the time scale
$\epsilon_n^{-1}$ set by the density~\cite{Parish_Levinsen_PRB2016}.
Moreover, recent experiments on the very-early-time dynamics of
quantum quenches in the Bose
gas~\cite{UniversalPrethermalDynamicsHadzibabic} suggest that a
similar situation holds for the unitary Bose gas.  To investigate this
further, one must generalize Eq.~\eqref{eq:Ansatz} to include
three-particle processes, but this is beyond the scope of this work
and will be the subject of future studies.

%
\subsection{Equations of motion}
\label{sec:eqmot}
As in Refs.~\cite{Sykes,CorsonBohn}, the equations of motion for the
variational parameters $c^{}_{0}(t)$ and $g^{}_{\vec{k}}(t)$ can be
derived from the Euler-Lagrange
equations~\cite{KramerLagrangian,CorsonBohn},
\begin{align*}
  \frac{d}{dt}\frac{\partial\mathcal{L}}{\partial\dot{c}^*_0}
  &=\frac{\partial\mathcal{L}}{\partial c^*_0} &
  \frac{d}{dt}\frac{\delta\mathcal{L}}{\delta\dot{g}^*_{\vec{k}}} &=
  \frac{\delta\mathcal{L}}{\delta g^*_{\vec{k}}} \; ,
\end{align*}
associated with the Lagrangian
\begin{equation*}
  \mathcal{L} = \frac{i}{2} \left[ \langle\psi(t)| \dot{\psi}(t) \rangle -
    \langle\dot{\psi}(t) |\psi(t) \rangle\right] - \langle
  \hat{H}\rangle\; .
\end{equation*}
When evaluating the contributions to $\langle \hat{H}\rangle$, we have
that $\langle \hat{H}^{}_{3} \rangle$ = 0 while
\begin{equation}
  \expval*{\hat{H}^{}_{4}} = \frac{U_{\Lambda}}{2V}
  \sump_{\vect{k},\vect{q}} \left[ 2 N_{\vec{k}} (t) N_{\vec{q}} (t) +
    x^{}_{\vec{k}} (t) x^{*}_{\vec{q}} (t) \right] \; .
\label{eq:h4con}
\end{equation}
Differently from~\cite{PostquenchRadzihovsky,KainLing}, we retain the
anomalous expectation values $x^{}_{\vec{k}}$ in
$\expval*{\hat{H}^{}_{4}}$. One arrives at the following equations of
motion:
\begin{align}
\label{eq:EqMotion_c0}
i\dot{c}^{}_0 &= U_\Lambda c^{}_0 n + \frac{U_\Lambda}{V}
\sump_{\vec{k}} \frac{c^{}_0|g^{}_{\vec{k}}|^2 + c^{*}_0
  g^{}_{\vec{k}}}{1 -
  |g^{}_{\vec{k}}|^2}\\
\nonumber
i\dot{g}^{}_{\vec{k}} &= 2\left[\epsilon^{}_{\vec{k}} + U_\Lambda n
\right] g^{}_{\vec{k}} + U_\Lambda  \big(2|c^{}_0|^2
  g^{}_{\vec{k}} + g_{\vec{k}}^2c^{*2}_0 \\
  &+ c^2_0\big) +\frac{U_\Lambda}{V} \sump_{\vec{q}}
  \frac{2g^{}_{\vec{k}} |g_{\vec{q}}|^2 + g^2_{\vec{k}}
    g^{*}_{\vec{q}} + g^{}_{\vec{q}}}{1 - |g_{\vec{q}}|^2} \; ,
\label{eq:EqMotion_gk}
\end{align}
that have to be solved for a set of initial conditions $c^{}_0(0)$ and
$g^{}_{\vec{k}}(0)$. Note that the total density
\begin{equation}
  n = n_0(t) + n_{ex} (t) = 
  |c_0(t)|^2 + \Frac{1}{V} \sump_{\vect{k}} N_{\vect{k}} (t)\; ,
\label{eq:densi}
\end{equation}
is conserved during the dynamics governed by
Eqs.~\eqref{eq:EqMotion_c0} and~\eqref{eq:EqMotion_gk}.

For instantaneous quenches $a_i \mapsto a_f$, the system is initially
($t=0$) characterized by the scattering length $a_i$, while at later
times $t>0$ it evolves with the interaction set by the final
scattering length $a_f$. Therefore, the equations of
motion~\eqref{eq:EqMotion_c0} and~\eqref{eq:EqMotion_gk} have to be
solved by using $U_\Lambda = U_{\Lambda,f} =
(1-2a_f\Lambda/\pi)^{-1}4\pi a_f/m$. The initial state is assumed to
be in equilibrium and thus can be found by minimizing
$\bra{\psi}\hat{H}-\mu\hat{N}\ket{\psi}$ with respect to
time-independent variational parameters $c_0 (0)$ and
$g_{\vec{k}}(0)$, where $U_\Lambda = U_{\Lambda,i}$ and the Lagrange
multiplier $\mu$ (i.e., the chemical potential) fixes the number of
particles. For example, for an initial weakly interacting gas $n a_i^3
\ll 1$, one has that
\begin{align}
  |c_0^{}(0)|^2 &= n \left(1 - \Frac{8}{3 \sqrt{\pi}} \sqrt{n a_i^3}
  \right) \simeq n \\
  g_{\vect{k}}^{} (0) &= \Frac{\sqrt{\epsilon_{\vect{k}}
      (\epsilon_{\vect{k}} + 2U_i n)} - (\epsilon_{\vect{k}} +
    nU_i)}{nU_i} \; ,
\end{align}
where $U_i = 4\pi a_i/m$. 

The last two terms in the equation for the condensate
dynamics~\eqref{eq:EqMotion_c0} self-consistently describe the
condensate depletion due to the scattering to finite momentum excited
states. The last three terms in the equation for the excited state
dynamics~\eqref{eq:EqMotion_gk} instead represent the correlations
between non-condensed atoms. As shown in App.~\ref{app:shall}, for
shallow quenches $na_{i,f}^3 \ll 1$, the dynamics is integrable, the
terms just described can be neglected, and the now simplified
Eqs.~\eqref{eq:Bogo_eqmotion_c0} and~\eqref{eq:Bogo_eqmotion_gk} can
be solved exactly.
However, for a generic quench from a weakly interacting gas $na_{i}^3
\ll 1$ to an arbitrary value of the final scattering length $a_f$, the
dynamics is not integrable and one has to carry out a numerical
analysis.

\begin{table}
\centering
\begin{tabular}{|c|c|c|c|c|c|c|}
  \hline
$a$  & $a k_n$ & $\xi$ & $\tau$ & $|E_B|^{-1} $ & $k_n^{-1}$ &
$\epsilon_n^{-1}$ \\
\hline\hline
$100a_0$ & $2.1\times 10^{-2}$ & $3$~$\mu$m & $9$~ms & {\color{gray}$20$~ns} & {\color{gray}$0.3$~$\mu$m} & {\color{gray}$80~\mu$s}\\
\hline
$1000a_0$ & $2.1\times 10^{-1}$ & $0.9$~$\mu$m & $0.9$~ms & $2$~$\mu$s & $0.3$~$\mu$m & $80~\mu$s\\
\hline
$60000a_0$ & $12$ & {\color{gray}$0.1$~$\mu$m} & {\color{gray}$15$~$\mu$s} &
{\color{gray}$6$~ms} & $0.3$~$\mu$m & $80~\mu$s\\ \hline
\end{tabular}
\caption{Dimensionless interaction strength $a k_n$ in units of
  $k_n = (6 \pi^2 n)^{1/3}$ and characteristic length and time
  scales for different values of scattering length $a$ (in units of
  the Bohr radius $a_0$) for a $^{39}$K gas with density
  $n=10^{12}$~cm$^{-3}$. We indicate in grey all those scales that
  we expect to be irrelevant in a specific regime. The value
  $a=60000a_0$ is the largest final scattering length we can
  numerically simulate.}
\label{tab:scale}
\end{table}
%
\subsection{System parameters}
\label{sec:param}
Before describing the results obtained from the numerical simulations,
we summarize here the system characteristic length and time scales
that we expect to be relevant for different values of the final
scattering length $a_f$.
For shallow quenches $na_{i,f}^3 \ll 1$, the early-time quasiparticle
dynamics can be derived by using the time-dependent Bogoliubov
approximation~\cite{NatuMuellerCorrelationsBogoliubov,Hung_2013} and
can thus be described is terms of the healing length and mean-field
time scales only~\cite{StringariPitaevskii}:
\begin{align}
  \xi &=\Frac{1}{\sqrt{8\pi a n}} & \tau &= \Frac{m}{4\pi a n} \; .
\label{eq:bogop}
\end{align}
In fact, it can be shown that, in this regime, the dynamics is
universal in terms of these two parameters, where $a$ is taken to be
the final scattering length $a_f$. Here, the Bose gas occupies the
metastable upper branch at positive energies, and the molecular bound
state is far below the continuum when $na_f^3 \ll 1$. As shown later
on, while atom-molecule coherent oscillations occur on a time scale
$|E_{\text{B}}|^{-1} \ll \tau$ (see Tab.~\ref{tab:scale}), their
amplitude is negligible in this limit since they rescale as $na_f^3
\ll 1$.

Increasing the value of the final scattering length $a_f$ beyond the
weakly interacting regime, we anticipate that all scales play a role
in the description of the dynamics --- typical values for a $^{39}$K
gas with density $n=10^{12}$~cm$^{-3}$ are listed in
Tab.~\ref{tab:scale}. However, in the unitary regime $a_f \to \infty$,
we expect the dynamics to recover a universal behaviour in terms of
the density scales only, i.e., in terms of the typical momentum and
energy scales:
\begin{align}
  k_n &=(6\pi^2n)^{1/3} & \epsilon_n &= \Frac{k_n^2}{2m} \; .
\label{eq:densc}
\end{align}
In particular, we will see that, in the unitary regime, the typical
growth time $\tau_{gr}$ of the number of excited particles in a given
momentum state $\vect{k}$ follows a universal scaling law when
rescaled by $\epsilon_n^{-1}$ and $k_n$, respectively. Here, all
other system scales become irrelevant in the very-early-time quench
dynamics of the gas.

Of course, the arguments above ignore the existence of Efimov trimers,
which introduce an additional length scale in the problem. Therefore,
one might wonder about the typical time scales at which three-body
correlations and losses become relevant in the quench dynamics.  It
was found in Ref.~\cite{Fletcher} that for a thermal Bose gas of
$^{39}$K atoms trapped in a harmonic potential and quenched to large
values of the scattering length ($na_f^3 \gg 1$), the three-body
contact $C_3$ was negligible in the early stages of the dynamics where
$t \lesssim 5\epsilon_n^{-1}$ --- note that the density in this
experiment was $n\simeq 2.8 \times 10^{13}$~cm$^{-3}$, so that
$\epsilon_n^{-1} \approx 9$~$\mu$s.
More recently, it was shown in
Ref.~\cite{UniversalPrethermalDynamicsHadzibabic} that, for a box
potential trap, interaction quenches to the unitary regime for a
degenerate $^{39}$K gas have a universal dynamics and are lossless up
to $t \simeq \epsilon_{n}^{-1}$ --- in that experiment a density of
$n=5.1\times 10^{12}$~cm$^{-3}$ was considered, and thus
$\epsilon_{n}^{-1} \simeq 27~\mu$s, while in the Tab.~\ref{tab:scale}
we have fixed $n=10^{12}$~cm$^{-3}$ and thus $\epsilon_{n}^{-1} \simeq
80~\mu$s.

\section{Results}
\label{sec:resul}
We describe here how the early-time quench dynamics of a Bose
condensate at zero temperature evolves as a function of the final
scattering length $a_f$ all the way towards unitarity. To this end, we
numerically solve the equations of motion~\eqref{eq:EqMotion_c0}
and~\eqref{eq:EqMotion_gk} for the specific case of instantaneous
quenches from a non-interacting gas ($a_i=0$, $c_0(0)=\sqrt{n}$ and
$g_{\vect{k}} (0)=0$) to a generic value $a_f$.
Since we are considering $s$-wave interactions, we can assume
spherical symmetry for the function $g_{\vec{k}} (t) = g_{k} (t)$. We
use a Gauss-Legendre quadrature routine on a grid of $M$ points in
$k$-space and integrate the equations of motion using a
5$^{\text{th}}$-order Runge-Kutta routine. We have checked our results
are converged with respect to the chosen time step.
The dynamics has therefore two regularization parameters, namely the
number of points $M$ on the Gauss-Legendre momentum grid and the
momentum cutoff $\Lambda$. As explained in App.~\ref{app:ConvLambdaN},
we have checked the convergence of our results with respect to both
parameters and extrapolated the results from the numerics to both $M
\to \infty$ and $\Lambda \to \infty$ limits.
Note that, in the limit $\Lambda \to \infty$, our results do not
depend on the density $n$ and the final scattering length $a_f$
separately; rather they depend on the dimensionless interaction
strength $a_f k_n$.

\begin{figure}
\includegraphics[width=0.42\textwidth]{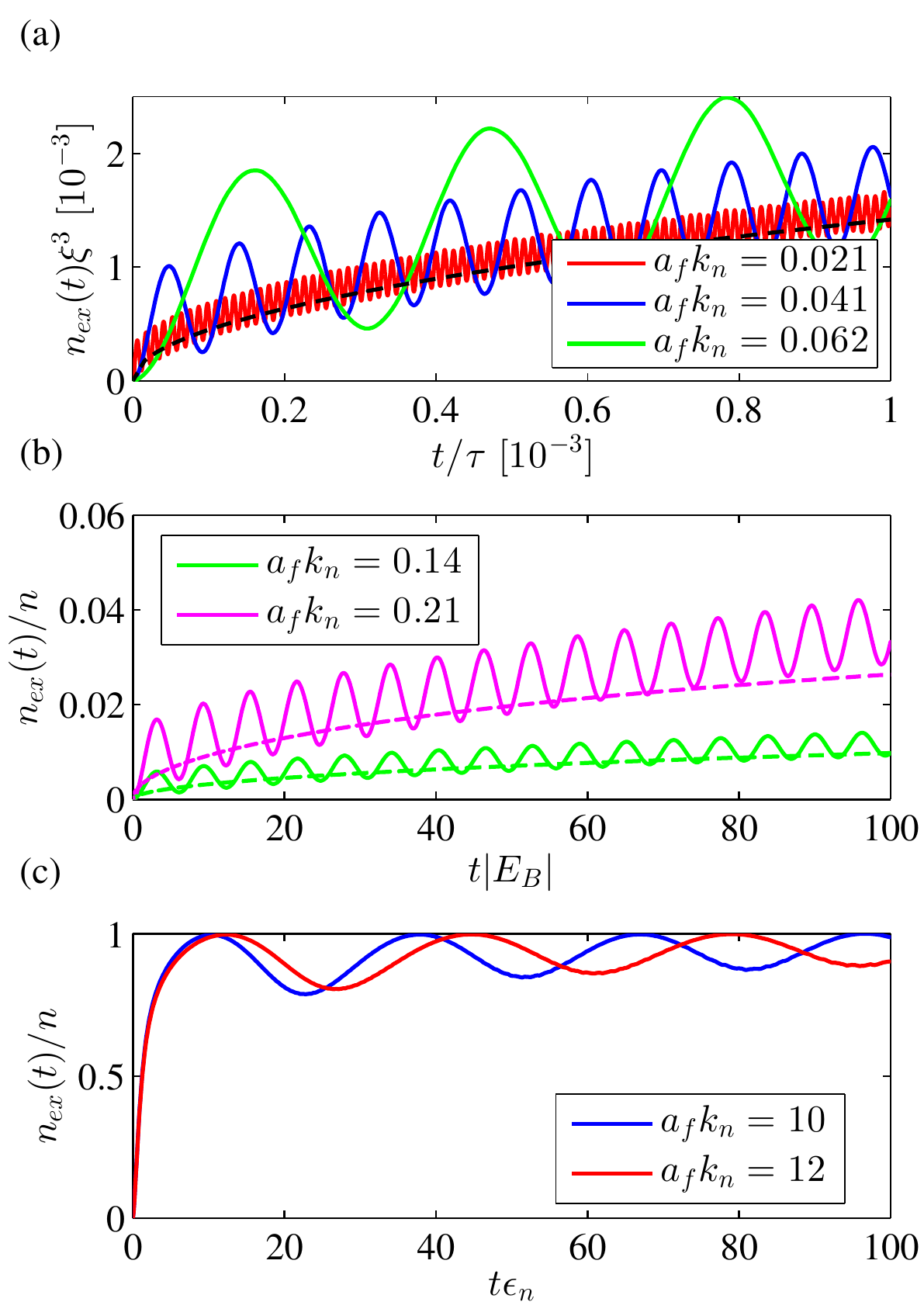}
\caption{Non-condensed density $n_{ex}(t)$ as a function of time after
  an instantaneous quench from a non-interacting gas $a_i=0$ to
  different values of the final scattering length $a_f$. In all
  panels, solid lines are the results of the numerical integration of
  the equations of motion~\eqref{eq:EqMotion_c0}
  and~\eqref{eq:EqMotion_gk}. (a) For shallow quenches, density and
  time are rescaled by the healing length $\xi$ and mean-field time
  $\tau$~\eqref{eq:bogop}, respectively. At intermediate values of
  $a_f$ (b), it is convenient to rescale the time by the molecular
  molecular bound state energy $|E_{\text{B}}|$, while, towards
  unitarity (c), by the energy scale fixed by the gas density
  $\epsilon_n$.  In panels (a) and (b), the dashed lines are obtained
  via the time-dependent Bogoliubov
  approximation~\cite{NatuMuellerCorrelationsBogoliubov} (see
  Eq.~\eqref{eq:BogNonCondFrac}).}
\label{fig:nex_vs_t}
\end{figure}
\begin{figure}
\includegraphics[width=0.5\textwidth]{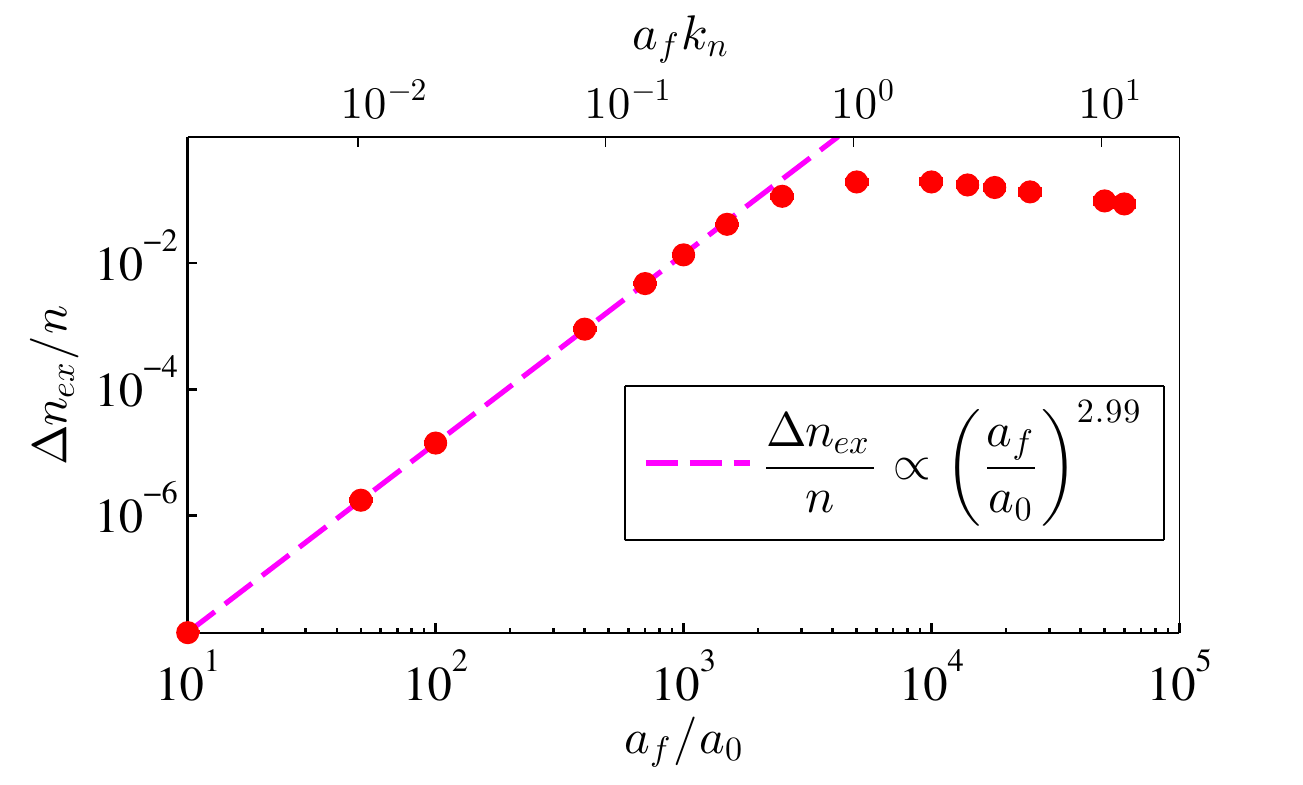}
\caption{Amplitude of coherent oscillations for the non-condensed
  fraction $\Delta n_{ex}$ as a function of $a_f k_n$ ($a_f$ in units
  of the Bohr radius $a_0$ is given for a density
  $n=10^{12}$~cm$^{-3}$).}
\label{fig:StdStateAmpOsc}
\end{figure}
%
\subsection{Non-condensed fraction}
\label{sec:nonco}
We plot in Fig.~\ref{fig:nex_vs_t} the density of particles in the
excited states $n_{ex} (t)$~\eqref{eq:densi} as a function of time
after an instantaneous quench from the non-interacting case $a_i=0$,
and for different values of the final scattering length $a_f$. It is
convenient to rescale both the non-condensed density $n_{ex}(t)$ and
the time $t$ by different scales depending on the different range of
final scattering lengths $a_f$ considered.
In the weakly interacting regime (panel (a)), we rescale density and
time using the healing length $\xi$ and mean-field time
$\tau$~\eqref{eq:bogop}, respectively. In this regime, the dynamics
obtained within the time-dependent Bogoliubov approximation ([black]
dashed line) is universal in these units (see App.~\ref{app:shall}),
i.e., it does not depend on any other scale of the problem. One can
easily solve Eq.~\eqref{eq:BogNonCondFrac} for the early-time
dynamics, obtaining a square-root behaviour for the initial increase
of $n_{ex}^{Bog}(t)$:
\begin{equation}
  n_{ex}^{Bog}(t) \xi^3 \Simiq_{t \ll \tau } \Frac{1}{4\pi^{3/2}} 
  \sqrt{\frac{t}{\tau}} \; .
\end{equation}

\begin{figure}
\centering
\includegraphics[width=0.5\textwidth]{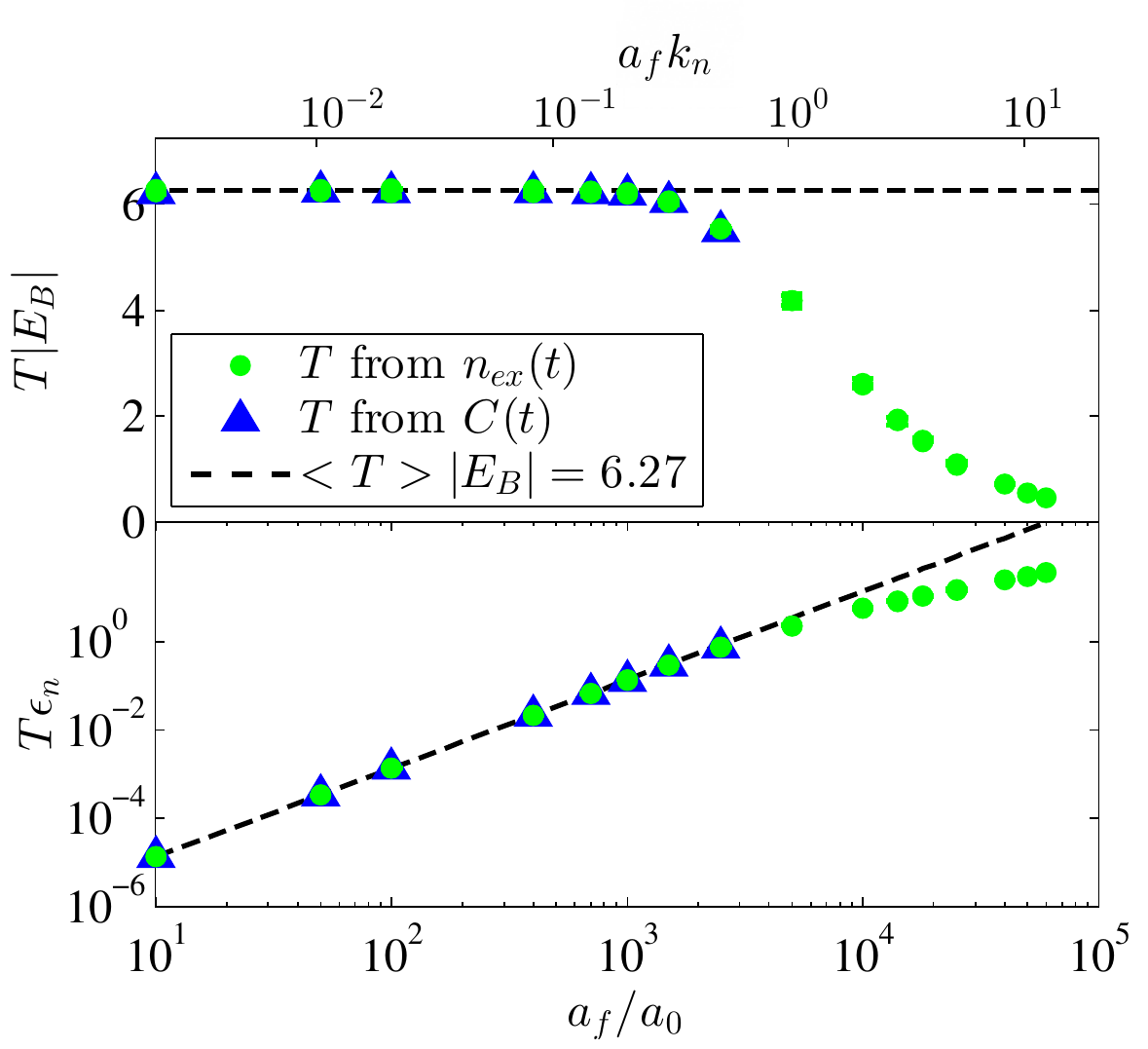}
\caption{Period $T$ of coherent oscillations extracted from the
  excited particle density $n_{ex}(t)$ ([green] dots) and from the
  Tan's contact $C(t)$ ([blue] triangles) as a function of $a_f k_n$
  ($a_f$ in units of the Bohr radius $a_0$ is given for a density
  $n=10^{12}$~cm$^{-3}$). Top panel: the period $T$ is plotted in
  units of the molecular binding energy $|E_B|$. Bottom panel: $T$ is
  plotted in units of the energy scale associated with the gas density
  $\epsilon_n$.}
\label{fig:PeriodVsAf}
\end{figure}

As Fig.~\ref{fig:nex_vs_t}(a) shows, the universal behaviour of
$n_{ex}^{Bog}(t)$ in units of $\xi$ and $\tau$ for shallow quenches is
only weakly modified by the inclusion of the condensate depletion as
well as the correlations between non-condensed atoms, which are both
contained in the equations of motion~\eqref{eq:EqMotion_c0}
and~\eqref{eq:EqMotion_gk}.
These terms, together with the renormalization of the interaction
strength~\eqref{eq:U_Lambda}, include a description of the molecular
bound state. As a consequence, coherent atom-molecule oscillations
appear in the density of excited particles, as already predicted by
Ref.~\cite{CorsonBohn}.  The oscillations are due to the reversible
transfer of pairs of atoms to the molecular bound
state~\cite{Donley_Nature_2002,Claussen_PRA_2003}.
The oscillation period $T$ is set by the molecular binding energy
$|E_{\text{B}}|$ and, as explained later, we find that $T \simeq
2\pi/|E_{\text{B}}|$ for a wide interval of $a_f$ values (see
Fig.~\ref{fig:PeriodVsAf}). For shallow quenches one has
$|E_{\text{B}}|^{-1} \ll \tau$ (see Tab.~\ref{tab:scale}) --- note
that, in panel (a) of Fig.~\ref{fig:nex_vs_t}, the oscillation period
increases with $a_f$ because the time $t$ is in units of the
mean-field time $\tau$ which decreases like $\propto a_f$ while
$|E_{\text{B}}|^{-1}$ increases like $\propto a_f^2$.
However, in this regime the amplitude of oscillations is negligible,
$\Delta n_{ex}/n \ll 1$, (see Fig.~\ref{fig:StdStateAmpOsc}) and thus
would be difficult to detect. In addition, far from the resonance,
the single-channel model employed in~\eqref{eq:hamil} may not provide
an accurate description of the interactions.

\begin{figure}
\centering
\includegraphics[width=0.45\textwidth]{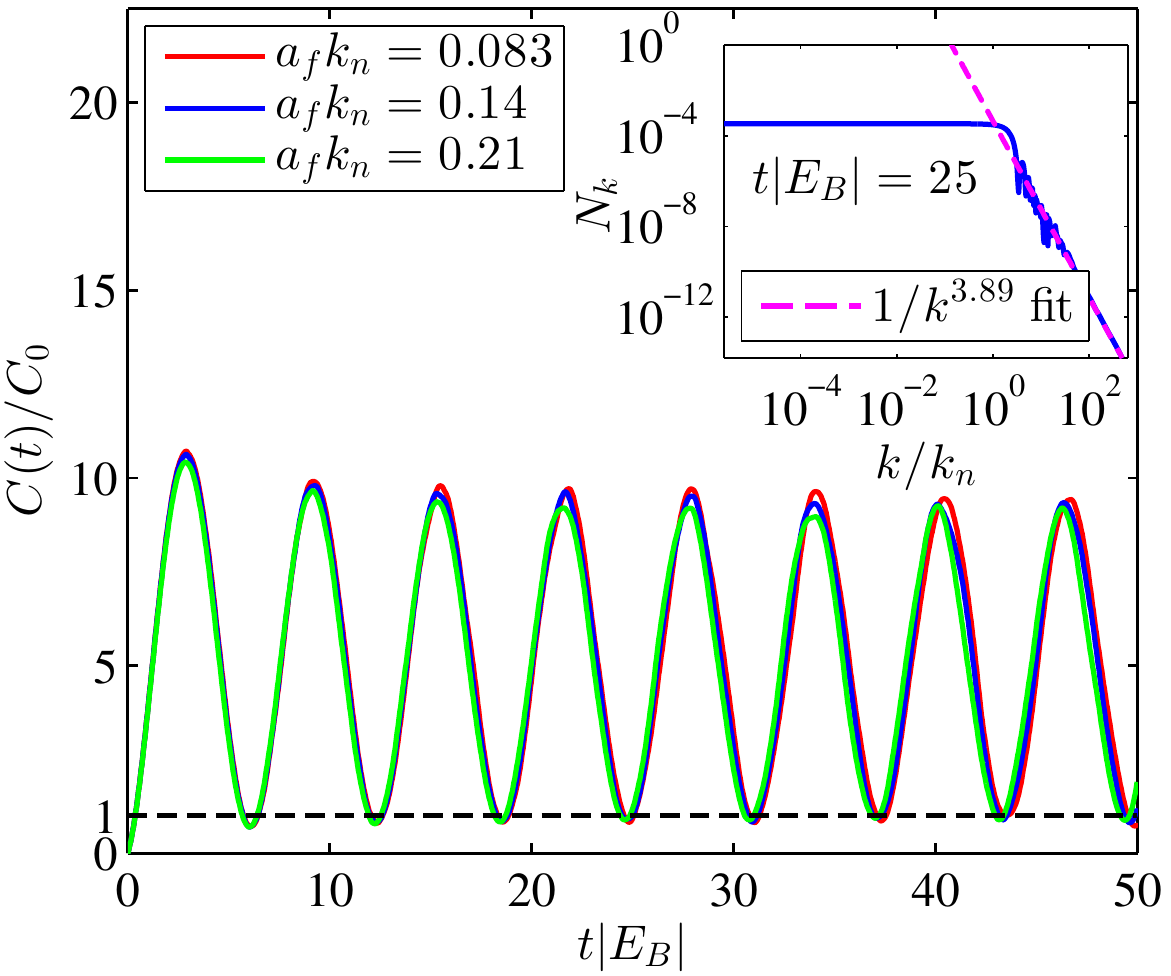}
\caption{Rescaled Tan's contact $C(t)/C_0$ ($C_0 = 16 \pi^2 a_f^2
  n^2$) for instantaneous quenches from a non-interacting gas $a_i=0$
  to different values of the final scattering length $a_f$. Inset:
  momentum distribution $N_{k} (t)$ at a fixed time $t$ and scattering
  length $a_f k_n=0.14$.}
\label{fig:Ct}
\end{figure}
By increasing the value of the final scattering length $a_f$, the
time-dependent Bogoliubov approximation eventually loses its validity,
primarily because it does not include self-consistently the depletion
of the condensate.
In Fig.~\ref{fig:nex_vs_t}(b) we plot the density of particles in the
excited states $n_{ex} (t)$ by rescaling the time by the molecular
binding energy.
In contrast to the case of shallow quenches, we can see that our
numerical results, at long enough times, start deviating from the ones
obtained within the time-dependent Bogoliubov approximation. This
deviation occurs on longer time scales (absolute units) for larger
values of $a_f$ --- in Fig.~\ref{fig:nex_vs_t}(b) the deviation occurs
on shorter time scales for larger values of $a_f$ because time is
measured in units of $|E_{\text{B}}|^{-1}$.
In addition, we find that the amplitude of the coherent atom-molecule
oscillations increases as $a_f^3$ (see Fig.~\ref{fig:StdStateAmpOsc})
and it reaches $\Delta n_{ex}/n \sim 0.12$ for $a_f k_n = 0.52$. Thus
we expect the oscillations to become relevant by increasing the value
of $a_f$.

We extract the amplitude of the oscillations by averaging over several
oscillations after the initial transitory behaviour, and using the
standard deviation as the uncertainty --- the error bar is too small
to be observed in Fig.~\ref{fig:StdStateAmpOsc}.
In addition, we find that the oscillation period is universal $T
\simeq 2\pi/|E_{\text{B}}|$ up to $a_fk_n \lesssim 0.21$, while the
medium starts to influence the dynamics by decreasing the period value
only for larger interaction strengths (top panel of
Fig.~\ref{fig:PeriodVsAf}).
Note that, for $a_f k_n \lesssim 0.21$, $|E_{\text{B}}|^{-1} \lesssim
2~\mu$s (see Tab.~\ref{tab:scale}). As commented at the end of
Sec.~\ref{sec:param}, recent
experiments~\cite{Fletcher,UniversalPrethermalDynamicsHadzibabic} have
shown that three-body processes and losses are negligible for $t
\lesssim \epsilon_{n}^{-1} \sim 80~\mu$s and thus there is an interval
in the early-time dynamics when we expect that it should be possible
to measure coherent atom-molecule oscillations before three-body
processes and losses start to affect the dynamics.

\begin{figure*}
\includegraphics[width=0.85\textwidth]{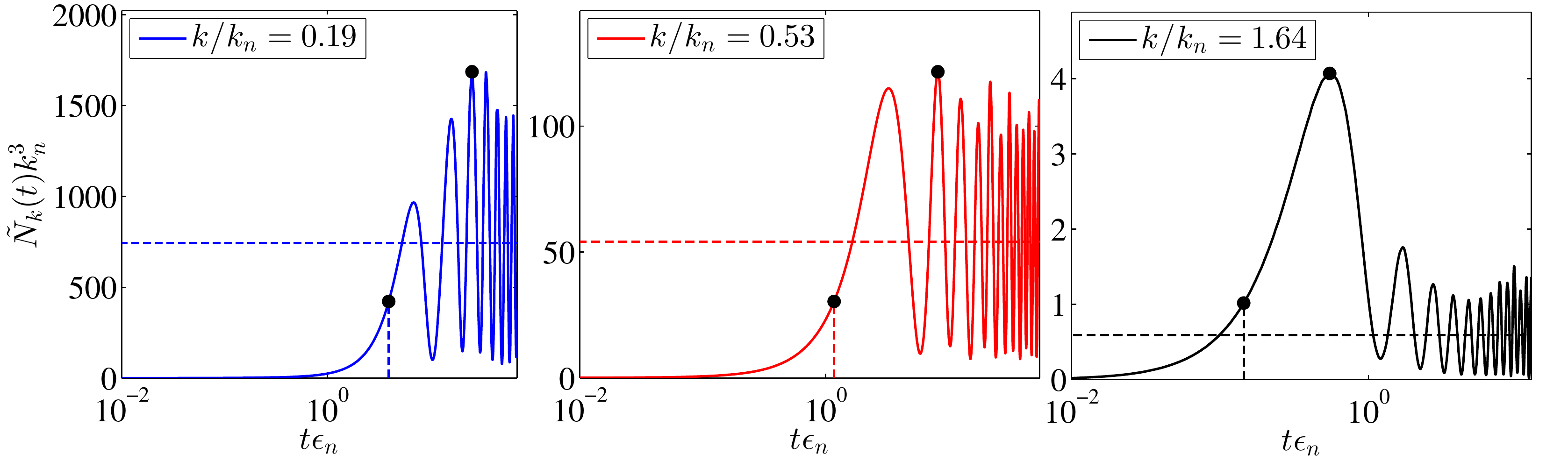}
\caption{Normalized momentum distribution $\tilde{N}_{\vect{k}} (t)$
  as a function of time $t$ after a diabatic quench to $a_f k_n=4.1$
  (corresponding to $a_f=20000a_0$ when $n=10^{12}$~cm$^{-3}$) for
  different value of the momentum $k$. The dashed horizontal line
  represents the value of the mean steady-state $\tilde{N}_{ss} (k)$
  reached after an initial quench transient. The dashed vertical line
  marks the value of the growth time $\tau_{gr} (k)$, defined as the
  time at which $\tilde{N}_{\vect{k}} (t)$ reaches $1/4$ of its
  maximum value ([black] dots) at a given value of $k$.}
\label{fig:Hadzibabic1}
\end{figure*}

Finally, we plot in Fig.~\ref{fig:nex_vs_t}(c) the excited state
density for the largest values of $a_f$ we can simulate. As explained
in App.~\ref{app:ConvLambdaN}, where we discuss the convergence of our
results with respect to the number of points $M$ on the Gauss-Legendre
momentum grid and the momentum cutoff $\Lambda$, our numerics suffers
a critical slowing down of the convergence with respect to both
parameters.
For this reason, $a_f k_n = 12.4$ ($a_f = 6\times 10^4 a_0$ and
$n=10^{12}$~cm$^{-3}$) is the largest value we can simulate to extract
converged information in a reasonable running time.
In this regime, we find that the excitation density very quickly
converges into a steady-state regime with a large average condensate
depletion of $\sim 96\%$. In addition, we find that the increase of
the period of oscillations $T$ with $a_f$ slows down for $a_f k_n
\gtrsim 0.21$.
In particular, in Fig.~\ref{fig:PeriodVsAf} we plot the period of the
coherent oscillations, either in units of $|E_{\text{B}}|$ (top panel)
or $\epsilon_n$ (bottom panel).
While the slope of $T\epsilon_n$ reduces sensibly, indicating the
approach to a universal regime, we cannot enter a universal regime for
the period where $T\epsilon_n \simeq \text{const}$.
At the same time, the period in this regime is $T > \epsilon_n^{-1}$
and on this time scale three-body events, heating and losses have been
shown to start playing a relevant role in the
experiments~\cite{Fletcher,UniversalPrethermalDynamicsHadzibabic},
making it difficult to measure the oscillations in $n_{ex} (t)$.
However, even though in the unitary regime we neither can disclose a
universal behaviour of the oscillation period nor expect this to be
accessible experimentally, as we will see later in
Sec.~\ref{sec:UniPreDy}, in the unitary regime we find a universal
scaling behaviour of the typical growth time $\tau_{gr}$ of the
momentum distribution which is a typical time of the very-early gas
dynamics ($\tau_{gr} \epsilon_n \lesssim 1$).

\subsection{Tan's contact}
\label{sec:Tans}
The momentum distribution of a quantum gas governed by an $s$-wave
contact interaction $a$ always decays for large momenta as
$k^{-4}$~\cite{TansContact}. The Tan's contact is thus defined as
\begin{equation}
  C = \lim_{k \to \infty} k^4 N_{\vec{k}} \; .
\label{eq:TansContact}
\end{equation}
$C$ measures the strength of two-body short-range correlations. For a
weakly interacting Bose gas, the Tan's contact can be easily derived
employing the Bogoliubov approximation and is given by $C_0 =
16\pi^2a^2n^2$~\cite{StringariPitaevskii}.
It is possible to show that the momentum distribution retains a
$k^{-4}$ large momentum tail also for quantum quenches, and thus one
can define an instantaneous Tan's contact
$C(t)$~\cite{CorsonBohn,PostquenchRadzihovsky}.

As shown in Fig.~\ref{fig:Ct}, we extract the time dependence of the
Tan's contact $C(t)$ after instantaneous quenches by fitting the tail
of the momentum distribution $N_{\vect{k}} (t)$ at fixed $t$ (see
inset) and plot it for several values of the scattering length $a_f$.
As for $n_{ex} (t)$, the Tan's contact also shows coherent
oscillations. The oscillation period coincides with that of $n_{ex}
(t)$ --- see Fig.~\ref{fig:PeriodVsAf}.
In addition, we show that there is a a large interval of values of
$a_f k_n \lesssim 0.21$ for which the oscillations are universal both
in amplitude (in units of $C_0 = 16 \pi^2 a_f^2 n^2$) as well as for
the period (in units of $|E_{\text{B}}|^{-1}$) and represent coherent
atom-molecule oscillations.
As commented in the case of $n_{ex}(t)$, because in this regime $T
\simeq 2\pi/|E_{\text{B}}| < \epsilon_n^{-1}$, we expect that the
coherent atom-molecule oscillations should be measurable by
considering the time dependence of the Tan's contact. However, for
larger values of $a_f$, eventually the period increases, reaching
$T>\epsilon_n^{-1}$, and it is thus on time-scales for which the
system is affected by three-body effects and losses.

Nevertheless, as discussed in the next section, our
Ansatz~\eqref{eq:Ansatz} is still able to describe the \emph{very
  early-time} dynamics, and it exposes a universal scaling behaviour
of the typical growth time of the momentum distribution, in agreement
with recent experiments~\cite{UniversalPrethermalDynamicsHadzibabic}.

\begin{figure}
\includegraphics[width=0.45\textwidth]{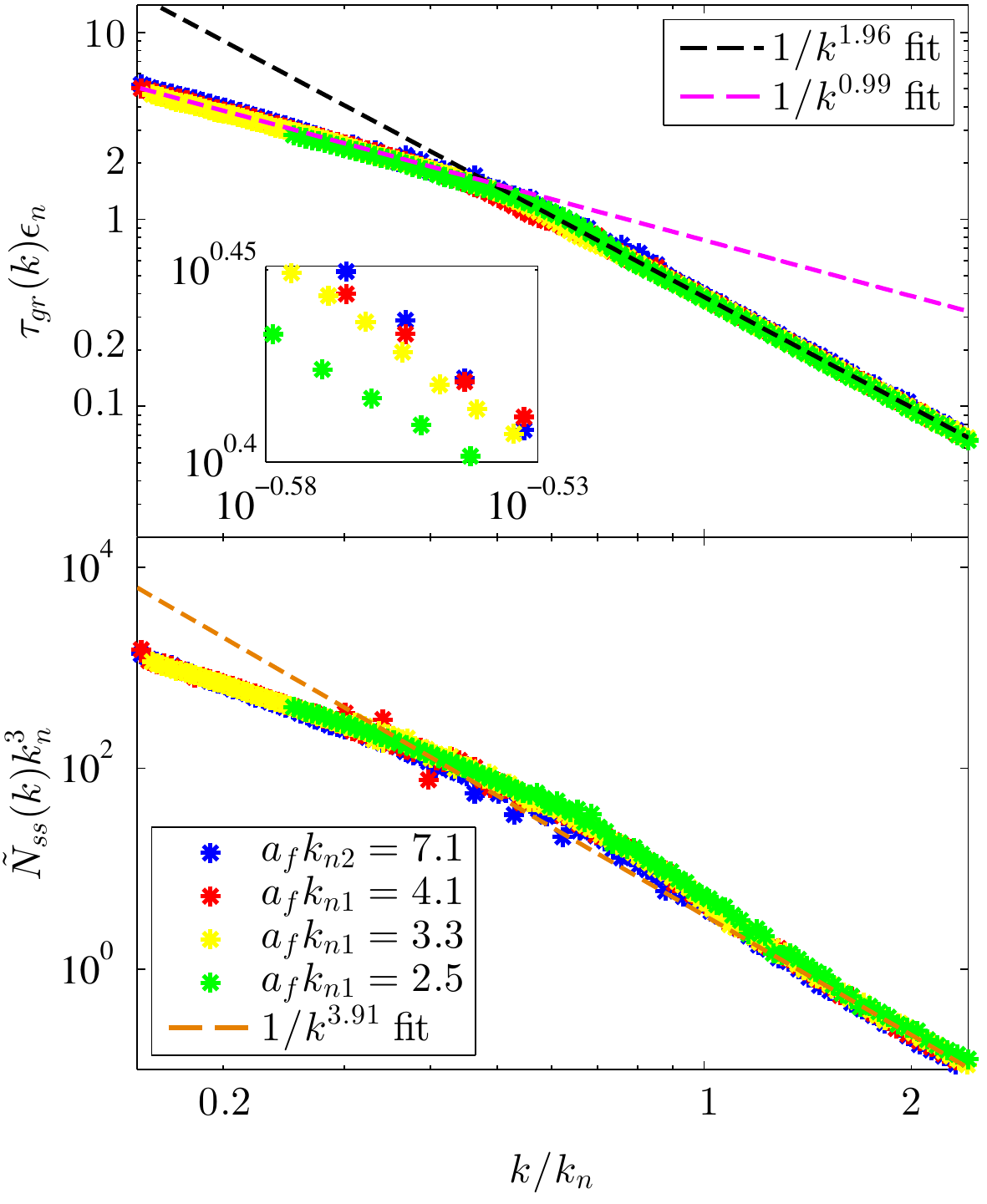}
\caption{Top panel: Growth time $\tau_{gr} (k)$ of the normalized
  momentum distribution $\tilde{N}_{\vect{k}} (t)$ (see
  Fig.~\ref{fig:Hadzibabic1}) as a function of momentum for different
  values of $a_f$ in the unitary regime and for two densities,
  $n_1=10^{12}$~cm$^{-3}$ and $n_2=5 \times 10^{12}$~cm$^{-3}$ (see
  legend in the bottom panel). Bottom panel: Mean steady-state value
  $\tilde{N}_{ss} (k)$ reached by the momentum distribution after an
  initial transient of the quench (see Fig.~\ref{fig:Hadzibabic1}) as
  a function of momentum for the same values of $a_f$ and densities as
  the top panel.}
\label{fig:Hadzibabic2}
\end{figure}
%
\subsection{Universal prethermal dynamics in the unitary regime}
\label{sec:UniPreDy}
Let us consider the time dependence of the momentum distribution
$N_{\vect{k}} (t)$ after a diabatic quench to $a_f$ in the unitary
regime. Similarly to
Ref.~\cite{UniversalPrethermalDynamicsHadzibabic}, we define the
normalized momentum distribution, as
\begin{align}
  \tilde{N}_{\vect{k}} (t) &= \Frac{N_{\vect{k}} (t)}{n_{ex} (t)} &
  \Frac{1}{V} \displaystyle \sump_{\vec{k}} \tilde{N}_{\vect{k}} (t)
  &= 1 \; .
\label{eq:normn}
\end{align}
In Fig.~\ref{fig:Hadzibabic1} we plot $\tilde{N}_{\vect{k}} (t)$ as a
function of time for different values of the momentum $k$. Immediately
after the quench, the momentum distribution grows rapidly as $\propto
t^{3/2}$.
We want to determine the typical growth time $\tau_{gr}(k)$ of the
momentum distribution at a fixed momentum. To this end, we identify,
at each fixed momentum $k$, the maximum value of $\tilde{N}_{\vect{k}}
(t)$ and define $\tau_{gr}(k)$ ([black] dashed vertical line) as the
time at which $\tilde{N}_{\vect{k}} (\tau_{gr})$ reaches a fixed
percentage of the maximum value.
We choose this to be $1/4$, but our results are independent of this
value, as long as $\tau_{gr}$ lies in the tail of the growing region
of $\tilde{N}_{\vect{k}} (t)$ that is not affected by the oscillatory
behaviour yet.
In fact, after the transient, the momentum distribution enters a
steady-state regime dominated by large coherent oscillations around a
steady state value $\tilde{N}_{ss} (k)$ ([black] dashed horizontal
line).
Note that these are not atom-molecule oscillations only because now
they involve the entire medium, as suggested by the fact that their
period $T$ moves away from $2\pi/|E_{\text{B}}|$ and starts to
saturate towards $\propto \epsilon_n^{-1}$.

The presence of large oscillations represents a substantial difference
from the experimental results of
Ref.~\cite{UniversalPrethermalDynamicsHadzibabic} where highly damped
oscillations are barely visible in the prethermal dynamics and where
the normalized momentum distribution quickly approaches a steady state
value after an almost monotonic initial growth.
For this reason, the authors of
Ref.~\cite{UniversalPrethermalDynamicsHadzibabic} can apply a sigmoid
fit to extract the growth time $\tau_{gr} (k)$ there defined as the
time at which the momentum distribution reaches $1/2$ of the
steady-state value $\tilde{N}_{ss} (k)$.
Even by averaging out the coherent oscillations in our results for
$\tilde{N}_{\vect{k}} (t)$, we cannot apply a sigmoid fit to our data
because, at large momenta, the growth towards the steady-state
prethermal regime is not monotonic (see the last panel of
Fig.~\ref{fig:Hadzibabic1}). For this reason we have applied the
criterion previously described to extract the growth time $\tau_{gr}
(k)$.

Nevertheless, by plotting in the top panel of
Fig.~\ref{fig:Hadzibabic2} the rescaled growth time $\tau_{gr}(k)
\epsilon_n$ as a function of the rescaled momentum $k/k_n$ we
disclose, in agreement with the
experiments~\cite{UniversalPrethermalDynamicsHadzibabic}, a universal
scaling behaviour which is independent of both the final scattering
length $a_f$ and the gas density $n$.
In particular, the value extracted for $\tau_{gr}(k)$ at different
values of $a_f k_n > 2.5$ and at two different gas densities, fall
onto the same universal curve. As
in~\cite{UniversalPrethermalDynamicsHadzibabic}, the fitting of our
data reveals the following universal scaling laws for small and large
momenta:
\begin{equation}
\begin{split}
  \tau_{gr}(k) \epsilon_n &\Simiq_{k < k_n}\Frac{k_n}{k}\\
  \tau_{gr}(k) \epsilon_n &\Simiq_{k >
    k_n}\left(\Frac{k_n}{k}\right)^2 \; .
\end{split}
\label{eq:uniun}
\end{equation}
We have checked that this result is independent of the specific
percentage of the maximum value of $\tilde{N}_{\vect{k}} (t)$ chosen
to define $\tau_{gr}(k)$ --- a different percentage just scales
rigidly all those curves up or down, leaving unchanged the universal
scaling behaviour.

Note that for shallow quenches one can also define a typical growth
time $\tau_{gr}^{Bog}(k)$ of the momentum distribution and a steady
state value $\tilde{N}_{ss}^{Bog} (k)$ that characterizes the
long-time dynamics of the momentum distribution. Starting from the
analytic expression~\eqref{eq:Bogog} for $g_{\vect{k}} (t)$ obtained
within the time-dependent Bogoliubov approximation, one can deduce the
following expressions:
\begin{align}
  \Frac{\tau^{Bog}_{gr}(k)}{\tau} &= \frac{\pi}{6 E_k\tau} &
  \Frac{\tilde{N}_{ss}^{Bog} (k)}{\xi^3} &= \frac{\sqrt{2^5}
    \pi}{(E_k\tau)^2} \; .
\end{align}
The dependence on the momentum $k$ of both quantities reveals
information about the Bogoliubov spectrum of quasi-particle
excitations of a weakly interacting Bose gas, $E_k\tau =
\sqrt{(k\xi)^2 [(k\xi)^2 + 2]}$. As in Eq.~\eqref{eq:uniun}, the
result for the growing time $\tau^{Bog}_{gr}(k)$ give the same
rescaling with momentum, i.e., at low momenta $k \xi \ll 1$,
$\tau^{Bog}_{gr}(k)/\tau \simeq (k\xi)^{-1}$, while in the opposite
regime $k \xi \gg 1$, $\tau^{Bog}_{gr}(k)/\tau \simeq
(k\xi)^{-2}$. This corresponds to the phonon and free particle regimes
of the Bogoliubov spectrum, respectively.
In the strongly interacting regime, we thus find the very same
universal regime but with the mean-field energy $\tau^{-1}$ replaced
by ``Fermi'' energy $\epsilon_n$ and the healing length $\xi$ by
$k_n^{-1}$. This thus indicates that the excitations spectrum is still
phonon-like for small momenta and free-like at higher momenta, but
with energy and momentum scales only determined by the gas density.
The agreement with the experiments on the universal scaling law of the
growth time $\tau_{gr}^{Bog}(k)$ indicates that the very-early-time
quench dynamics into the unitary regime, when a stationary prethermal
regime has not been reached yet, is dominated by excitations of atoms
out of the condensate in pairs only.

We plot in the bottom panel of Fig.~\ref{fig:Hadzibabic2} the mean
steady-state value $\tilde{N}_{ss} (k)$ reached by the momentum
distribution after the initial transient after the initial growth as a
function of momentum. Like in the experiments, we also find a
universal behaviour of this quantity in units of $k_n$.
However, in contrast with experiments for which this quantity follows
a single exponential law, our calculations reveal a crossover to a
power-law $k^{-4}$ decay for large values of momenta. This is expected
for a dissipationless quantum gas governed by an $s$-wave contact
interaction, and leads to the extraction of the Tan's contact.
This indicates that, for quenches into the unitary regime, once the
very early-time growth and transient dynamics of the momentum
distribution has passed and the system enters the prethermal region,
our dissipationless description which leads to strong coherent
oscillations around a steady-state value is incomplete.

\section{Conclusions}
\label{sec:conc}
We have analyzed the crossover from shallow to deep instantaneous
interaction quenches in the early-time dynamics of a degenerate Bose
gas at zero temperature.
We have employed a time-dependent Nozi\`eres-Saint James variational
formalism, which self-consistently describes the excitation of
particles out of the condensate in pairs only. We have modelled
short-range atom interactions close to a Feshbach resonance using a
single-channel model, which admits a molecular bound state on the
repulsive side of the resonance.
The coupled dynamics between the condensate and the excited states
includes the condensate depletion and correlations between
non-condensed atoms.

In agreement with previous studies~\cite{Sykes,CorsonBohn}, we have
found coherent atom-molecule oscillations in both the density of
excited particles and the Tan's contact, and we have characterized
them as a function of the final scattering length $a_f$.
For shallow quenches, the oscillations have a negligible amplitude and
the dynamics is well captured by a time-dependent Bogoliubov
theory~\cite{NatuMuellerCorrelationsBogoliubov,Hung_2013}, involving
the mean-field time and the healing length.  However, at intermediate
values of the final scattering length $a_fk_n \lesssim 0.21$, we find
a universal regime for atom-molecule oscillations, where the period is
only determined by the molecular binding energy, $T \simeq
2\pi/|E_{\text{B}}|$, and the amplitude of oscillations is not
negligible.
We expect such oscillations to be visible in experiments, since recent
experiments on $^{39}$K have shown that three-body processes and
losses are negligible for $t \lesssim \epsilon_{n}^{-1} \sim
80~\mu$s~\cite{UniversalPrethermalDynamicsHadzibabic}, while
$|E_{\text{B}}|^{-1} \lesssim 2~\mu$s.

We have pushed our analysis into the unitary regime and we can
simulate the dynamics up to $a_f k_n = 12.4$. Here we find that the
growth time of the momentum distribution $\tau_{gr}(k)$, which
characterizes the very-early-time quench dynamics, depends on the
momentum $k$ with a universal scaling law that is governed by density
only. In particular, we can deduce that, at very short times after the
quench, the excitations are Bogoliubov modes with the mean-field
energy $\tau^{-1}$ replaced by $\epsilon_n$ and the healing length
$\xi$ by $k_n^{-1}$.
During these very-early times after the quench, where the momentum
distribution has not yet entered a prethermal regime, the agreement
between our results and the experiments indicates that higher-order
incoherent processes do not affect the postquench behaviour at short
times and thus, the employment of a time-dependent Nozi\`eres-Saint
James variational formalism is adequate in this case.
However, the very strong damping of the coherent oscillations observed
in experiments when the quench dynamics enters the prethermal regime,
indicates that damping and dissipation mechanisms should eventually be
taken into account.
This is further confirmed by the observation of an exponential
behaviour of the mean steady-state value $\tilde{N}_{ss} (k)$ reached
by the momentum distribution in the prethermal
regime~\cite{UniversalPrethermalDynamicsHadzibabic}, while instead our
model predicts a typical power-law $k^{-4}$ decay for large values of
momenta.

One way to include damping and dissipation would be to consider a
phenomenological model, as in Ref.~\cite{Rancon_Levin_PRA_2014}, where
the system is coupled to an external bath, thus allowing energy to
dissipate.
Alternatively, one could consider the next order term in the
time-dependent variational Ansatz~\eqref{eq:Ansatz} which allows the
excitations of particles out of the condensate in triplets. The
inclusion of this term would allow one to incorporate the effects of
Beliaev decay, Landau
scattering~\cite{PrethermalisationThermalisationCrossoverIacopo} and
potentially even Efimov physics. This will be the subject of future
studies.

\acknowledgments We acknowledge useful discussions with I. Carusotto
and L. Tarruell.
F.M.M. acknowledges financial support from the Ministerio de
Econom\'ia y Competitividad (MINECO), projects
No.~MAT2014-53119-C2-1-R and No.~MAT2017-83772-R.  M.M.P. acknowledges
support from the Australian Research Council Centre of Excellence in
Future Low-Energy Electronics Technologies (CE170100039).
%

\appendix

\section{Shallow quenches}
\label{app:shall}
For shallow interaction quenches, $n a_{i,f}^3 \ll 1$, the dynamics is
integrable and one can solve exactly~\eqref{eq:EqMotion_c0}
and~\eqref{eq:EqMotion_gk}, recovering the results of
Refs.~\cite{NatuMuellerCorrelationsBogoliubov,Hung_2013}. In this
limit, the contribution from $\langle \hat{H}^{}_4 \rangle
$~\eqref{eq:h4con} can be neglected and one can replace the bare
interactions with inverse zero-energy T matrix, giving the interaction
strength $U_{i,f} = 4\pi a_{i,f}/m$. Moreover one can assume the
condensate depletion is negligible, $|c_0(t)|^2 \simeq n$, and thus
the equation of motions can be simplified to:
\begin{align}
\label{eq:Bogo_eqmotion_c0}
  i\dot{c}^{}_0 &\simeq U_f c^{}_0 n\\
  i\dot{g}^{}_{\vec{k}} &= 2\left[\epsilon^{}_{\vec{k}} + 2U_f n
  \right] g^{}_{\vec{k}} + U_f \left(g_{\vec{k}}^2c^{*2}_0 +
    c^2_0\right) \; .
\label{eq:Bogo_eqmotion_gk}
\end{align}
It is easy to show that these equations are solved exactly by 
\begin{align}
  c^{}_0(t) &= \sqrt{n} e^{-i U_f n t}\\
  g^{}_{\vec{k}}(t) &= \bar{g}^{}_{\vec{k}}(t) e^{-2 i U_f n t}\\
  \bar{g}^{}_\vec{k}(t) &= \frac{E_{\vec{k}f} g_{\vec{k}}(0) -
    i\tan(E_{\vec{k}f}t) [U_f n + \xi_{\vec{k}f}
    g_{\vec{k}}(0)]}{E_{\vec{k}f} + i\tan(E_{\vec{k}f}t)
    [\xi_{\vec{k}f} + U_f ng_{\vec{k}}(0)]} \; ,
\label{eq:Bogog}
\end{align}
where $E_{\vec{k}f} = \sqrt{\epsilon_{\vect{k}} (\epsilon_{\vect{k}} +
  2 U_f n)}$ is the quasi-particle excitation spectrum and
$\xi_{\vec{k}f} = \epsilon_{\vect{k}} + U_f n$.

This coincides with the result obtained in
Refs.~\cite{NatuMuellerCorrelationsBogoliubov,Hung_2013} by using a
time-dependent Bogoliubov approximation, where one considers the
Heisenberg equations of motion for the particle operator
$\hat{a}_{\vect{k}}^{}(t)$,
\begin{align*}
  i \Frac{d \hat{a}_{\vect{k}}^{}}{dt} = [\hat{a}_{\vect{k}}^{},
  \sum_{\vec{k}} \epsilon_{\vec{k}}^{} \hat{a}_{\vect{k}}^{\dagger}
  \hat{a}_{\vect{k}}^{} + \hat{H}^{}_2 ] \; .
\end{align*}
This equation is solved in terms of the Bogoliubov parameters,
$\hat{a}_{\vect{k}}^{} = u_{\vect{k}} (t) \hat{b}_{\vect{k}}^{} +
v_{\vect{k}}^* (t) \hat{b}_{-\vect{k}}^{\dag}$, where $|u_{\vect{k}}
(t)|^2 - |v_{\vect{k}} (t)|^2= 1 $, giving:
\begin{equation*}
  i \Frac{d }{dt} \begin{pmatrix} u_{\vect{k}} (t) \\ v_{\vect{k}}
    (t) \end{pmatrix} = \begin{pmatrix} \xi_{\vect{k}f} &
    U_f n \\ -U_f n &
    -\xi_{\vect{k}f} \end{pmatrix} \begin{pmatrix} u_{\vect{k}} (t) \\
    v_{\vect{k}}
    (t) \end{pmatrix}\; .
\end{equation*}
It is easy to show that these coupled equations are solved by
\begin{align}
  \bar{g}_{\vect{k}} (t) &= \Frac{v_{\vect{k}}^* (t)}{u_{\vect{k}}^*
    (t)} & |u_{\vect{k}} (t)| &= \Frac{1}{\sqrt{1 -
      |\bar{g}_{\vect{k}}(t)|^2}} \; ,
\end{align}
and one recovers, e.g. for the fraction of particles in the excited
states, the result of \cite{NatuMuellerCorrelationsBogoliubov}:
\begin{multline}
n_{ex}^{Bog}(t)\xi^3=\frac{1}{2\pi^2}\int_{0}^{\tilde{\Lambda}}\tilde{k}^2d\tilde{k}\Biggl[
\frac{|g_k(0)|^2}{1-|g_k(0)|^2}
\\
-\left(\frac{a_i}{a}-1\right)
\frac{\sin^2\left(\sqrt{\tilde{k}^2(\tilde{k}^2+2)}\tau\right)}{(\tilde{k}^2+2)\sqrt{\tilde{k}^2(\tilde{k}^2+2a_i/a)}}
\Biggr]
\; .
\label{eq:BogNonCondFrac}
\end{multline}
\begin{figure}
\centering
\includegraphics[width=0.4\textwidth]{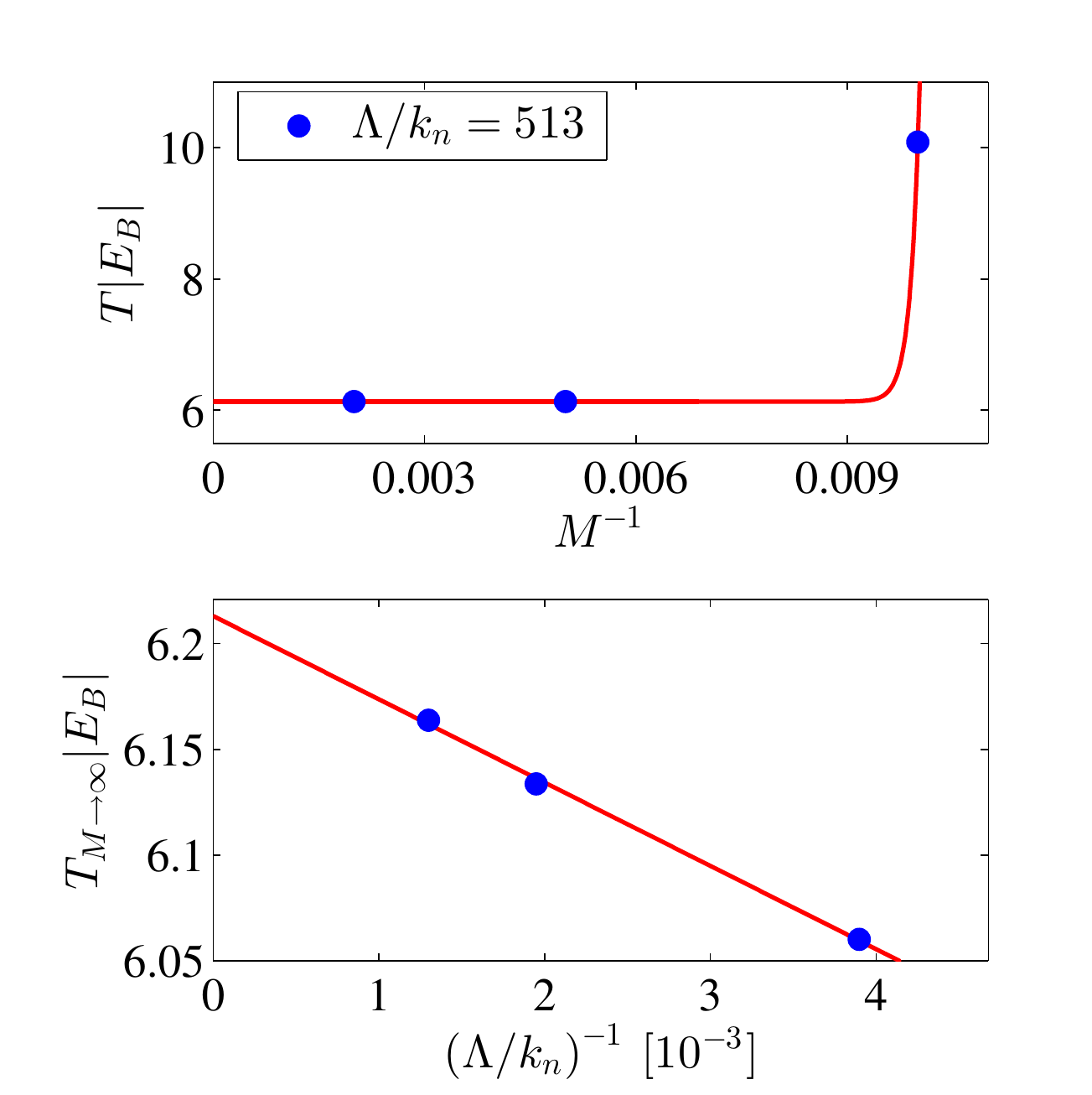}
\caption{Two-step extrapolation process of the period of oscillations
  $T$ of the non-condensed density $n_{ex}(t)$
  (Fig.~\ref{fig:nex_vs_t}) with respect to the two system
  regularisation parameters, $M$ and $\Lambda$. The period $T$ ([blue]
  dots) is evaluated, at fixed values of $M$ and $\Lambda$, by
  averaging over several oscillations of $n_{ex}(t)$, with an error
  given by the standard deviation (not observed in these plots). Top
  panel: $T$ is plotted as a function of $M^{-1}$ for a fixed value of
  $\Lambda/k_n=513$. The $M \to \infty$ value $T_{M \to \infty}$ is
  extracted via an exponential fit ([red] solid line). Bottom panel:
  Plot of the extracted period $T_{M \to \infty}$ with respect to
  $\Lambda^{-1}$ ([blue] dots). The extrapolated value $T_{M \to
    \infty, \Lambda \to \infty}$ is obtained via a linear fit ([red]
  solid line). We have fixed $n=10^{12}$~cm$^{-3}$ and $a_f=1000a_0$
  ($a_f k_n = 0.21$).}
\label{fig:extrp}
\end{figure}
%
\section{Convergence of the dynamics}
\label{app:ConvLambdaN}
We show here the convergence of our results with respect to the number
of points $M$ on the Gauss-Legendre momentum grid and the momentum
cutoff $\Lambda$. Note that, as the dynamics requires in principle one
regularization parameter only, we could send $\Lambda \to \infty$ and
use only the number of points $M$ to regularize
Eq.~\eqref{eq:U_Lambda}. However, we could not obtain results
converged in time applying this procedure. For this reason, we have
fixed both $M$ and $\Lambda$ in our numerics and checked the
convergence with respect to both parameters.

We show here the extrapolation procedure followed to extract the $M
\to \infty$ and $\Lambda \to \infty$ results for the oscillation
period $T$ of the non-condensed density $n_{ex}(t)$. Convergence with
respect to both parameters has been checked for all data reported in
the main text. We have found that the dynamics converges exponentially
fast with respect to the number of Gauss-Legendre points $M$ used for
the quadrature, while it only has a linear dependence on the cutoff
$\Lambda$ (for the range of $\Lambda$ values computationally
accessible).
Fig.~\ref{fig:extrp} shows the dependence of the period of
oscillations with respect to both $M$ (top panel) and $\Lambda$
(bottom panel). Once we have extracted $T_{M\to\infty}$ for different
values of $\Lambda$, we can extract the final value $T_{M \to \infty,
  \Lambda \to \infty}$ reported in Fig.~\ref{fig:PeriodVsAf}.
The numerical calculations are limited by a critical slowing down of
the convergence with respect to the regularization parameters for both
very small and very large values of $a_f$.
%

%

\end{document}